\documentclass[10pt]{article}
\usepackage{graphicx}
\usepackage{dcolumn}
\usepackage{bm}
\usepackage{amsmath}
\usepackage{amssymb}
\usepackage{hhline}
\usepackage[normalem]{ulem}

\usepackage{amsfonts}

\usepackage{graphicx}
\usepackage{dcolumn}
\usepackage{bm}
\usepackage{amsmath}
\usepackage{amssymb}
\usepackage{revsymb4-1}
\usepackage{hhline}

\begin{document}

\begin{center}
\textbf{\Large Artificial gravity effect on spin-polarized exciton-polaritons}\\

E. S. Sedov$^{1,2,*}$, A. V. Kavokin$^{3,1,4+}$\\

$^1${School of Physics and Astronomy, University of Southampton, SO17 1NJ Southampton, United Kingdom}

$^2${Department of Physics and Applied Mathematics, Vladimir State University named after A. G. and N. G. Stoletovs, Gorky str. 87, Vladimir 600000, Russia}

$^3${CNR-SPIN, Viale del Politecnico 1, I-00133, Rome, Italy}

$^4${Spin Optics Laboratory, St. Petersburg State University, Ul'anovskaya 1, Peterhof, St. Petersburg 198504, Russia}\\

$^{*}$ {evgeny\_sedov@mail.ru}, $^{+}$ {a.kavokin@soton.ac.uk}
\end{center}

\begin{abstract}
The pseudospin dynamics of long-living exciton-polaritons in a wedged 2D cavity has been studied theoretically accounting for the external magnetic field effect.
The cavity width variation plays the role of the artificial gravitational force acting on a massive particle: exciton-polariton.
A semi-classical model of the spin-polarization dynamics of ballistically propagating exciton-polaritons has been developed.
It has been shown that for the specific choice of the magnetic field magnitude and the initial polariton wave vector the polariton polarization vector tends to an attractor on the Poincar\'{e} sphere.
Based on this effect, the switching of the polariton polarization in the ballistic regime has been demonstrated.
The self-interference of the polariton field emitted by a point-like source has been shown to induce the formation of  interference patterns.
\end{abstract}

\section*{Introduction}


Exciton-polaritons in 2D semiconductor cavity structures being an equipollent combination of cavity photons and elementary matter excitations (excitons) and exhibiting properties of both their constituents represent a fascinating object for study both from the fundamental point of view and for applications in opto-electronics.
Until recently, one of the key constraints for observation of long-range polariton propagation was the extremely short polariton lifetime which in the best case wouldn't exceed few tens of picoseconds~\cite{Nature4434092006,Science33211672011}.
Since exciton-polaritons are composite quasiparticles arising due to the strong coupling of photons and excitons, they have a finite lifetime that is governed by lifetime of photons in the cavity~\cite{ContemporaryPhysics521312011}.
In recent papers~\cite{PRX30410152013,PRB882353142013} an increase of the polariton lifetime up to of the order of 100~ps due to the use of specially designed 2D cavities with ultra-high quality factors has been reported.
In these structures, polaritons can move away from the excitation spot by the distances of the order of a few millimetres in a cavity plane.
The possibility of generation of a long-range coherent polariton flow in high-finesse microcavities is very promising for realization of ultrafast optoelectronic devices. 


Given the possibility of long-living polaritons excitation, a subject of the control of their spatial dynamics arises.
One of the convenient approaches to manipulation of exciton-polariton dynamical characteristics is through the generation of an external potential of a required configuration.
This idea is widely used in optics, e.g. for observation of optical Bloch oscillations in waveguide arrays, where the gradient of optical properties is provided by the linear variation of the effective index of individual guides~\cite{PhysRevLett8347561999} or by the temperature gradient applied to the thermo-optic polymer arrays~\cite{PRL8347521999}.
In this context, the idea of polariton acceleration in a cavity with the gradient of the cavity thickness has been discussed in \cite{PRB64081303R2001} and then developed in~\cite{Optica21}.
Reduction of the cavity width leads to a positive shift of the lower  dispersion branch (LB) of polaritons.
Since the latter is concave, to conserve the energy fixed, LB polaritons are forced to reduce their wave vector component in the propagation direction.
Their group velocity in the same direction decreases as a result.
As it has been shown in~\cite{Optica21}, in samples where the cavity width decreases linearly with one of the in-plane coordinates polaritons propagate on parabolic trajectories.
Herewith, contribution of a cavity thickness variation can be described as an effect of an external constant force analogous to the gravitational force acting on massive particles.

The effect of the gravitational force on quantum systems including atomic systems and Bose-Einstein condensates has been widely discussed within last few decades~\cite{PRL824681999,AmJPhys677761999,PRL8335771999}.
The bouncing of quantum wave packets on a hard surface under the effect of the gravitational force has been considered theoretically in Ref.~\cite{AmJPhys677761999}. 
The experimental work~\cite{PRL8335771999} is devoted to the observation of the bouncing of the Bose-Einstein condensate of $^{87}$Rb atoms off a mirror induced by a repulsive dipole potential under the gravity.
In parallel, the concept of simulation of gravitational effects using artificial systems that reproduce particular characteristics of the actual gravitational system in the laboratory has been developed.
This concept is known as \textit{the analogue gravity}~\cite{GenRelandGrav3417192002,LivingRevRelativ1432011}.
The analogue models are mostly based on various condensed-matter systems including flowing fluids~\cite{PRL4613511981,PRL1060213022011} and superfluids~\cite{PZETF688331988,PRD580640211998}, uniform~\cite{Science31913672008,NaturePhotonics16532007,OpticsExpress15145602007,PRL1021804022009,NaturePhysics118722015} and stratified~\cite{PRL1142374022015,JETPL104622016} semiconductor and dielectric media.
The shaping of light belonging to the short-wavelength spectral range of the continuum by the effect of the gravity-like force created by accelerating solitons has been demonstrated in~\cite{NaturePhotonics16532007}.
In Ref.~\cite{OpticsExpress15145602007} authors have theoretically predicted the optical effect analogous to the mentioned above quantum bouncing known for cold atoms.
This effect manifests itself as a multiple reflection of a short optical pulse propagating in an optical fiber on a refractive index variation created by a co-propagating continuously decelerated soliton. 
Experimental observation of bouncing of light in curved optical waveguides has been reported in~\cite{PRL1021804022009}.
In Ref.~\cite{NaturePhysics118722015} the effect of the self-induced ``artificial gravity'' on the interaction of optical wave packets in the Newton-Schr\"{o}dinger system has been investigated.
The nonlinear effects are responsible for the appearance of the ``gravitational'' potential in that system.
The effects of gravitational lensing, tidal forces and gravitational redshift and blueshift have been emulated.
The nonlinear optical Newton-Schr\"{o}dinger system has been investigated as the analogue of a rotating boson star evolution in Ref.~\cite{NatureCommun7134922016}.

Due to the specificity of the polariton dynamics under the effect of ``artificial gravity'',  polaritons have been observed to slow down until a complete stop and then reverse their propagation direction~\cite{Optica21}.
The authors of Ref.~\cite{Optica21} refer to such type of motion as ``slow reflection''.
This effect has been proposed in~\cite{PRB852351022012} as a tool for suppression of the light reflection by the edge of the sample wich is essential for diminishing the parasitic effects caused by interference of polaritons propagating towards the edge with the reflected ones.
In contrast with the normal reflection, in the case of ``slow reflection'' the phase of propagating polaritons does not change abruptly at the turning point but varies continuously.
The latter point is essential for understanding of the non-trivial interference of the direct and reflected polariton waves, as we show below. 

Exciton-polaritons are widely discussed as potential candidates for all-optical information transport, storage and processing because of their non-trivial polarization properties.
The cavity polariton polarization is mostly dependent on two factors that are the long-range electron-hole exchange interaction~\cite{PRB47157761993} and the polarization splitting of the photonic cavity modes~\cite{AKSpEffExcPolCond2012}.
In particular, the TE-TM splitting of the cavity eigenmodes leads to the splitting of linearly polarized polariton dispersion branches that induces the precession of the polarization vector of propagating polaritons~\cite{SuperlatandMicrostr413132007,PRB5950821999}.
One can describe the polariton polarization using the pseudospin formalism~\cite{KavokinBaumbergMicrocavities2007,PRL109036404,PRL920174012004}.
The polariton TE-TM splitting effect can be described as the polariton pseudospin precession around an effective  magnetic field applied in the cavity plane.
The non-trivial polariton pseudospin dynamics has been revealed in a wide range of experimentally observed efects including the polarization multistability~\cite{PRL982364012007}, spin switching~\cite{NatPhoton43612010}, optical spin Hall effect~\cite{PRL109036404}, etc.
In the presence of the external magnetic field $\mathbf{B}$, the polariton pseudospin dynamics becomes even more complicated. 
For the field applied in the Faraday geometry ($\mathbf{B}$ is normal to the cavity plane and is collinear with the incident light wave vector), the exciton-polariton energy band structure is enriched with the Zeeman splitting of right- and left-circularly polarized polariton states~\cite{SciRep6200912016,PRB781653232008,PRB880353112013}.

In this paper, we consider the pseudospin dynamics of long-living polaritons in a wedged 2D microcavity under the influence of the ``gravitational'' force produced by the cavity width gradient in the presence of the external magnetic field applied normally to the cavity plane.
The paper is organized as following.
In the following section we discuss the model describing the polariton polarization dynamics within the pseudospin formalism.
The next section presents the study of a particular case of polariton propagation in the gradient direction.
The last section is devoted to a general case of the polariton propagation oblique to the effective ``gravitational'' field.
In conclusion, we address the importance of polariton self-interference and polarization effects in the ``slow reflection'' regime.

\section*{Ballistic Motion of Exciton-Polaritons in Graded Microcavities}

\begin{figure}[t]
\includegraphics[width= 0.55\textwidth]{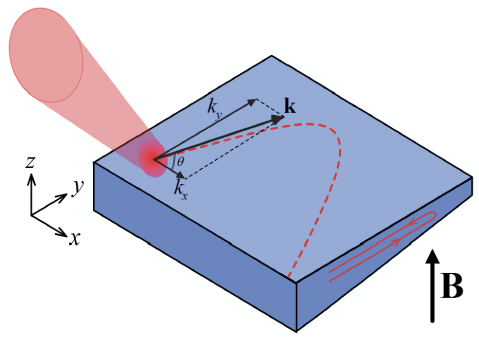}
\caption{\label{GeneralScheme} (Color online) The schematic of the considered problem.
The sample represents a 2D electrodynamical microcavity with embedded ensemble of semiconductor quantum wells.
The width of the cavity in $z$ direction gradually decreases along $y$ axis. 
Polaritons are injected in the sample by a laser beam inclined to the cavity plane.
By manipulation the angle of inclination, one possible to control the polariton kinetic energy in the sample in $xy$ plane.
An external magnetic field~$\mathbf{B}$ is applied to the system normally to the cavity plane.}
\end{figure}

The considered geometry of the problem is schematically shown in Fig.~\ref{GeneralScheme}.
It basically coincides with that studied experimentally in Refs.~\cite{PRX30410152013,Optica21}.
The sample represents a semiconductor microcavity with embedded quantum wells that are holders of excitons.
The cavity is characterized by a linear variation of the thickness in $y$-direction that causes a polariton energy gradient in the same direction.
The exciton-polaritons are supposed to be excited resonantly by the external laser beam.
We shell consider both the regime of excitation by ultrashort laser pulses and the regime of continuous wave pumping.
The initial quantum state of generated polaritons, their polarization, energy and wave vector are supposed to be set by the exciting laser.

Propagation of the polaritons in microcavities can be described by the following effective Hamiltonian written in the basis of right and left circular polarizations:
\begin{equation}
\label{TotHam}
\widehat{H} = \widehat{T} +\widehat{V} + \widehat{H} _{\text{LT}}  + \widehat{H} _{\text{M}},
\end{equation}
where
$\widehat{T} = \frac{\hbar^2 \widehat{k}^2}{2 m^{*}} \widehat{I}$ and
$\widehat{V} = \hbar \beta y \widehat{I}$ are the kinetic energy and the potential energy operators.
$\widehat{ \mathbf{k} } = ( \widehat{k}_x,\widehat{k}_x) = (-i\partial _{x}, -i \partial _{y})$ is the polariton momentum operator, 
$m^{*}$ is the effective mass of polaritons,
$\widehat{I}$ is the $2 \times 2$ unit matrix.
The Hamiltonian~\eqref{TotHam} acts upon a two-component wave function  $ \mathbf{\Psi} (t,\mathbf{r}) = \left( \Psi _{+} (t,\mathbf{r}), \Psi _{-} (t,\mathbf{r}) \right) ^{\text{T}}$, where ``$+$'' and ``$-$'' correspond to the $+1$ and $-1$ projections of the polariton spin on the $z$ axis.
The impact of the ``gravitational'' force $F = - \hbar \beta$ induced by the sample thickness gradient is taken into account by introducing the linear variation of the polariton potential energy in $y$-direction.

The Hamiltonians $\widehat{H} _{\text{LT}} $ and $ \widehat{H} _{\text{M}}$ describe the longitudinal-transverse splitting of linear polarizations and the effect of the normal to the cavity plane external magnetic field that induces the Zeeman splitting of circular polarizations, respectively.
They are given by
\begin{equation}
\label{LTandRealMagnFieldHams}
\widehat{H} _{\text{LT}}=
\frac{\hbar}{2}
\begin{bmatrix}
0 & \left( \widehat{\Omega} _{x} - i \widehat{\Omega} _{y} \right) \\
\left(\widehat{\Omega} _{x} + i \widehat{\Omega} _{y} \right) & 0
\end{bmatrix},
\qquad
\widehat{H} _{\text{M}}=
\frac{\hbar}{2}
\begin{bmatrix}
\widehat{\Omega} _z & 0 \\
0 & - \widehat{\Omega} _z
\end{bmatrix}.
\end{equation}

The operators $\widehat{\Omega} _{x}$ and $\widehat{\Omega} _{y}$ in~\eqref{LTandRealMagnFieldHams} represent components of the effective  magnetic field (gauge field) operator $\widehat{\mathbf{\Omega}} = \left( \widehat{\Omega} _x,\widehat{\Omega} _y, \widehat{\Omega} _z \right)$ that affects the polariton propagation:
\begin{equation}
\label{OmxyDef}
\widehat{\Omega} _{x}=\Delta _{\text{LT}} \left( \widehat{k}_{x}^{2} - \widehat{k}_{y}^{2}\right),
\qquad
\widehat{\Omega} _{y}= 2\Delta _{\text{LT}} \widehat{k}_{x} \widehat{k}_{y},
\end{equation}
where $\Delta _{\text{LT}}$ is the TE-TM splitting constant.

In the absence of external magnetic fields, $\widehat{\Omega} _{z} = 0$.
The switching of the magnetic field $\mathbf{B}$ in normal to the cavity plane direction ($B_{x,y} = 0$) removes the degeneracy in $z$ direction that leads to the appearance of the third component of the operator~$\widehat{\mathbf{\Omega}}$:
\begin{equation}
\label{OmzDef}
\widehat{\Omega} _{z}=\frac{\mu _{\text{B}} g B}{\hbar },
\end{equation}
where $\mu _{\text{B}}$ is the Bohr magneton, $g$ is the polariton g-factor and $B \equiv |\mathbf{B}|$.

In the simplest model case, the dynamics of a single polariton as well as one of a polariton wave packet can be considered as a particle-like dynamics of its ``center of mass'' along the trajectory  $\mathbf{r}_c (t) = (x_{c} (t), y_{c} (t))$.
The center of mass is characterized by the wave vector $\mathbf{k}_c (t) = (k_{cx} (t), k_{cy} (t))$.
Let us introduce a polariton wave function in a center-of-mass frame~\cite{ProgrTheorPhys867831991,MeisterArnoldMollArxiv2017,PRE640566022001}:
\begin{equation}
\label{ShiftedWFDef}
\mathbf{\Psi} (t,\mathbf{r}) = e^{i \chi (t)}  \mathbf{\Phi} (t, \mathbf{r} - \mathbf{r}_c),
\end{equation}
where 
$\chi(t)$ is the global phase that arises due to the classical action calculated along the center-of-mass trajectory $\mathbf{r}_c$.
The time-dependent vectors $\mathbf{r}_c$ and $\mathbf{k}_c$ obey the classical equations of motion
\begin{equation}
\label{ClassTrajectEqns}
\partial _{t} \mathbf{r}_c = \frac{\hbar}{m^{*}} \mathbf{k}_c, \qquad
\partial _{t} \mathbf{k} _c = -\beta \mathbf{e}_y
\end{equation}
corresponing to the classical Lagrangian density $\mathcal{L} = \frac{m}{2} \left( \partial _t \mathbf{r} _c\right)^2 - \hbar \beta y_c$;
$\mathbf{e}_y$ is the unit vector in the direction of $y$.
Here we neglected by the small effect of the LT-splitting on the center-of-mass motion.
For the initial conditions fully determined by the excitation light properties and taken as $\mathbf{r}_{0} = (0,0)$ and $ \mathbf{k}_{0} = (k_{x0},k_{y0})$ the solutions of~\eqref{ClassTrajectEqns} are easily found as
$x_{c}= \left. \hbar k_{x0} t \right/ m^{*}$,
$y_{c}= \left. \hbar k_{y0} t \right/ m^{*} - \left. \hbar \beta t^2 \right/ 2 m^{*}$,
$k_{cx} = k_{x0}$ and
$k_{cy} = k_{y0}-\beta t$.

For a single particle ballistically propagating along the center-of-mass trajectory we finally arrive at the set of coupled equations for the wave function amplitude $\mathbf{\Phi} (t) = \left( \Phi _{-}(t), \Phi _{+} (t) \right)$:
\begin{equation}
i \partial _t \Phi _{\pm} =
\pm \frac{1}{2} \Omega _{cz} \Phi _{\pm} +
\frac{1}{2} \left( \Omega _{cx} \mp i \Omega _{cy} \right) \Phi _{\mp},
\end{equation}
where $\Omega _{cx, cy, cz}$ characterize the components of the effective magnetic field $\mathbf{\Omega} _c$ affecting the  polariton propagation.
$\Omega _{cx, cy, cz}$ are described by Eqs.~\eqref{OmxyDef}--\eqref{OmzDef} where we substitute $\widehat{k}_{x,y} \rightarrow k _{cx, cy}$.
The global phase is found as~\cite{ProgrTheorPhys867831991,MeisterArnoldMollArxiv2017}
\begin{equation}
\label{GlobalPhaseDef}
\chi (t) = \int _{0} ^{t} \left( \mathcal{L} (t') - \frac{1}{2} \partial _{t'} (\mathbf{k}_c (t') \mathbf{r}_c (t')) \right) d t' =  - \frac{\hbar \beta t^2}{12 m^{*}} \left(3 k_{y0} - \beta t \right).
\end{equation}

Considering the propagation of a single cavity polariton, we can fully characterize its polarization (pseudospin) state by the Stokes vector $\mathbf{S} = \left(S_x,S_y,S_z \right)$, where $S_{x,y,z}$ are the intensities of linear (collinear with the original coordinate basis axes components and diagonal/antidiagonal ones) and circular polarization components, respectively~\cite{KavokinBaumbergMicrocavities2007}.
The vectors $\mathbf{\Phi}$ and $\mathbf{S}$ are linked between themselves by the following set of expressions:
\begin{equation}
\label{StokesDefThroughEPM}
S_{x}=\frac{1}{2} \left( \Phi_{+} \Phi _{-}^{*} + \Phi _{+} ^{*} \Phi _{-} \right), 
\,
S_{y}=\frac{i}{2} \left( \Phi_{+} \Phi _{-}^{*} - \Phi _{+} ^{*} \Phi _{-} \right), 
\,
S_{z}= \frac{1}{2} \left( |\Phi _{+}|^2 - |\Phi _{-}|^2\right).
\end{equation}

The squared absolute value of the polariton wave function is dependent on  $S_0 = \sqrt{S_x^2 + S_y^2+ S_z^2}$.
For the fully polarized wave $S_0$ remains unchanged during the polariton propagation.
Using this fact, in further discussions we express $S_{x,y,z}$ in the units of~$S_{0}$; $|S_{x,y,z}| \le 1$.
Finally we obtain the following equation characterizing  the dynamics of the Stokes vector of the particle propagating along the center-of-mass trajectory~\cite{PRL109036404}:
\begin{equation}
\label{StokesCompOnTime}
\frac{d\mathbf{S}}{dt}= \mathbf{\Omega}_{c} \times \mathbf{S},
\end{equation}

It is important to underline that in contrast with the problem considered in Ref.~\cite{PRL109036404}, the absolute value and the orientation of the vector $\mathbf{\Omega}_c$ acting on a specific ballistically propagating polariton becomes time- and coordinate-dependent due to the ``effective gravity''.
The polarization components oscillate with a frequency $\Omega_c = |\mathbf{\Omega}_c|$ that is changing with the particle propagation.

Below we list the parameters used for the numerical modelling.
Following~\cite{Optica21}, we consider polaritons possessing the effective mass of $m^{*} = 5 \times 10^{-5} m_{e}$ with $m_{e}$ being the vacuum electron mass.
The LT-splitting constant is taken $\Delta _{\text{LT}} = 11.9 \, \mu\text{eV} \times \mu\text{m} ^{2}$.
We consider a sample which provides the ``gravitational'' force $F=-10.5 \, \mu\text{eV} / \mu\text{m}$.
Polariton g-factor is $g=0.1$.
The initial wave number $k_0$ can be tuned by changing the angle of incidence of the pump.


\section*{Polariton Propagation in the Direction of the Thickness Gradient}

\subsection*{Single Polariton Propagation}

Let us first consider polaritons propagating in the ``gravitational'' force vector direction ($y$-direction).
To control the polarization state of polaritons, we have three external parameters.
The first one is the ``gravitational'' force strength governed by the parameter~$\beta$.
In the absence of external magnetic fields ($B = 0$) the ``gravitational'' force doesn't affect $k_{x}$ component of the wave vector thus it remains unchanged during the polariton propagation.
In the considered case $k_{x} = 0$, according to Eq.~\eqref{OmxyDef} we have $\Omega _{cy} = \Omega _{cz} = 0$,
so that, Eq.~\eqref{StokesCompOnTime} yields $\partial _t S_{x}  = 0$.
Thus for the initial conditions $S_{y0}=S_{z0}=0$ and $S_{x0} = \pm 1$ the linear polarization is conserved in the course of the polariton propagation and the Stokes vector points to $(\pm 1,0,0)$ on the Poincar\'{e} sphere.

For the initial conditions $S_{y0} \ne  0 $ or/and $S_{z0} \ne 0 $ and $|S_{x0}| < 1 $ the parameter $S_{x}$ remains constant while $S_{y,z}$ components oscillate in the range  $(-\sqrt{1-S_{x0} ^{2}},\sqrt{1-S_{x0} ^{2}})$.
The solution of Eq.~\eqref{StokesCompOnTime} can be easily found analytically as
\begin{equation}
\label{SimplestSolutions}
S_x = S_{x0},\,
S_y = S_{y0} \cos \left[ \kappa(t) \right] + S_{z0} \sin \left[ \kappa(t) \right], \,
S_z = S_{z0} \cos \left[ \kappa(t) \right] - S_{y0} \sin \left[ \kappa(t) \right],
\end{equation}
where the propagation factor is $\kappa(t) = \Delta _{\text{LT}} t \left[ k_{y0}^2 - k_{y0} \beta t + \left.\beta ^2 t ^2 \right/ 3 \right] $.
The energy transfer between the circular and the diagonal linear polarization modes is clearly seen.
The polariton Stokes vector describes a circular trajectory on the Poincar\'{e} sphere at $S_{x} = S_{x0}$ in this case.

Another polarization dynamics control parameter is the initial wave vector $\mathbf{k}_0$ in the $xy$ plane.
Slightly changing  the value of $k_0$ one can enhance or diminish the impact of the ``gravitational'' force on the dynamics of the polarization components.

The third control parameter is an external magnetic field perpendicular to the cavity plane.
It makes $S_{x}$ time-dependent and allows for manipulating the output polarization when keeping $k$ unchanged.

\begin{figure}[t]
\includegraphics[width=1.\textwidth]{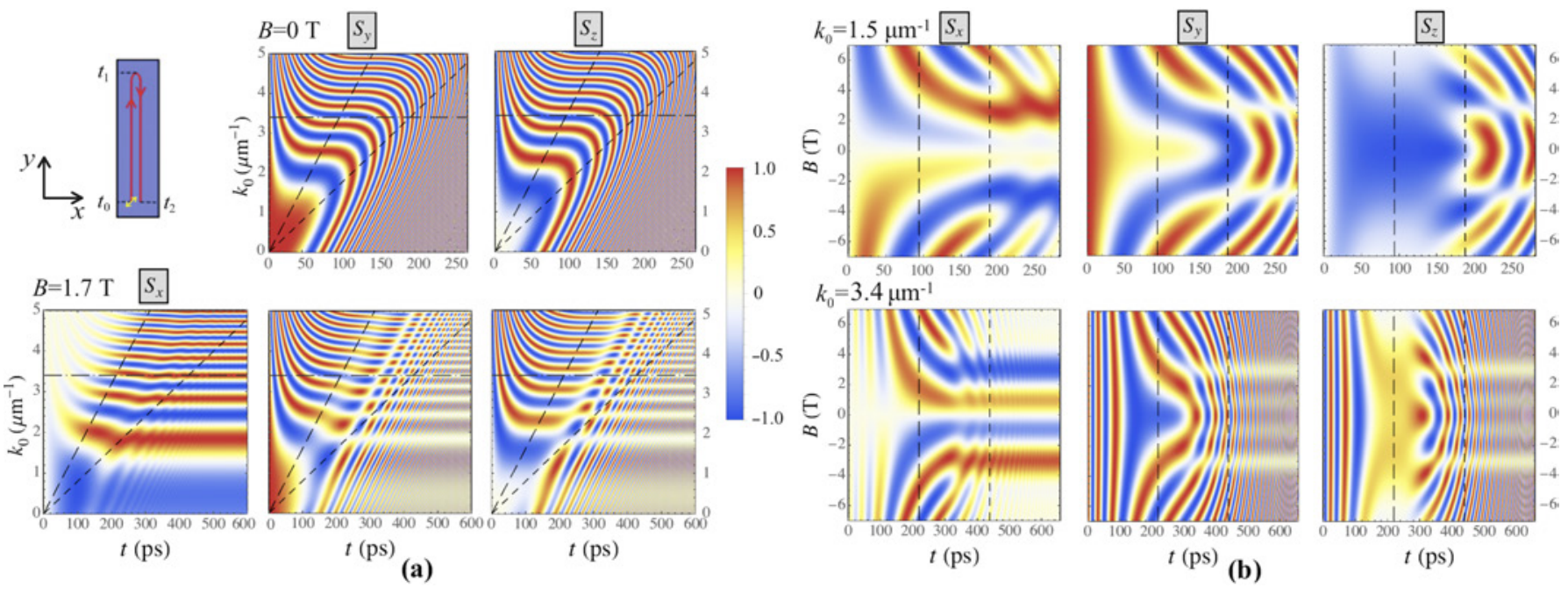}
\caption{\label{ColorMapsLin} (Color online)
The dynamics of the polariton polarization components $S_{x,y,z}$ in the case of polariton propagation along the gradient of the thickness as a function of the initial wave number $k_0$ (a) for $B=0$~T (the upper panels) and $1.7$~T (the lower panels), and as a function of the magnetic field $B$ (b) for $k_0=1.5 \, \mu \text{m}^{-1}$ (the upper panels) and $3.4 \, \mu \text{m}^{-1}$ (the lower panels)
The left upper panel in (a) shows schematically the polariton trajectory.
The initial polarization is taken diagonal linear with the initial conditions $S_{y0} = 1$, $S_{x,z0} = 0$.
The parameters used for this calculation are given in the text of the paper.
}
\end{figure}

\begin{figure}[t]
\includegraphics[width= 1.\textwidth]{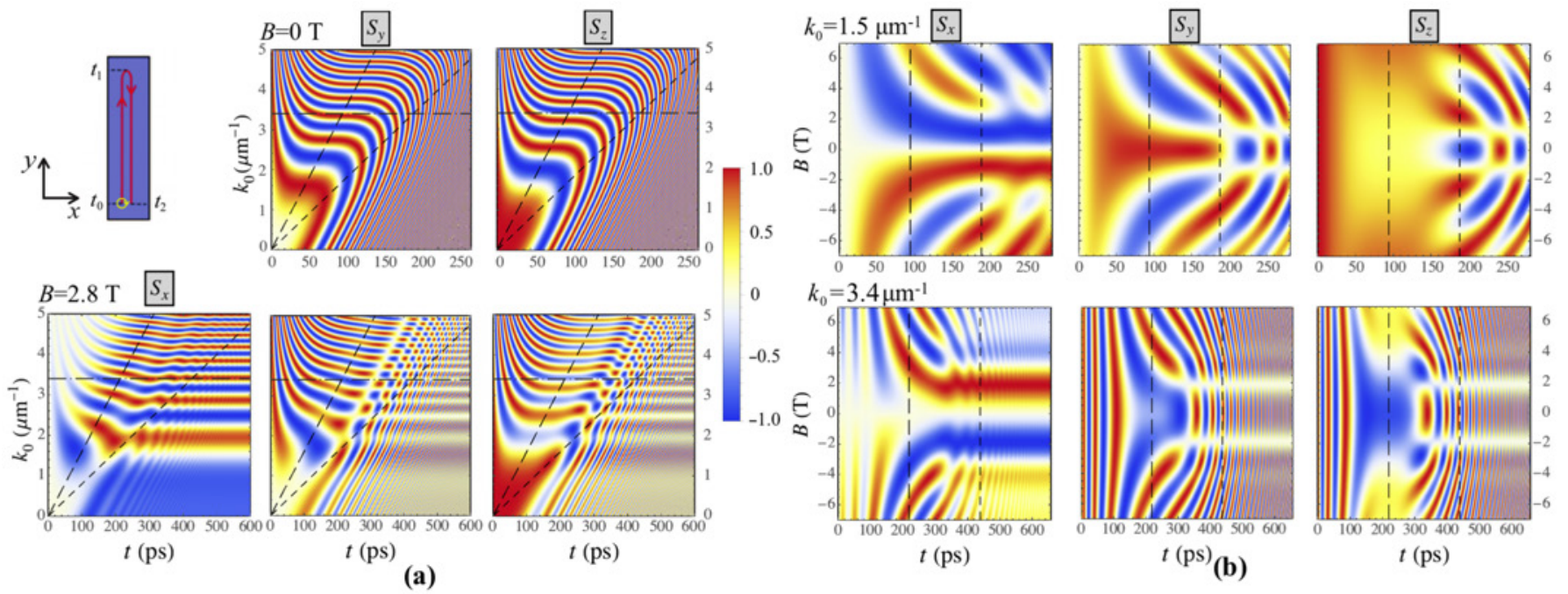}
\caption{\label{ColorMapsCirc} (Color online) 
The dynamics of the polariton polarization components $S_{x,y,z}$ in the case of polariton propagation along the gradient of the thickness as a function of the initial wave number $k_0$ (a) for $B=0$~T (the upper panels) and $2.8$~T (the lower panels), and as a function of the magnetic field $B$ (b) for $k_0=1.5 \, \mu \text{m}^{-1}$ (the upper panels) and $3.4 \, \mu \text{m}^{-1}$ (the lower panels).
The left upper panel in (a) shows schematically the polariton trajectory.
The initial polarization is taken circular with the initial conditions $S_{z0} = 1$, $S_{x,y0} = 0$.
The parameters used for this calculation are the same as in Fig.~\ref{ColorMapsLin}.}
\end{figure}

The color maps in Fig.~\ref{ColorMapsLin} illustrate the dynamics of the polariton polarization components $S_{x,y,z}$ as functions of the initial wave number $k_0$ for a number of values of the external magnetic field magnitude~$B$ (Fig.~\ref{ColorMapsLin}(a)) and on the value of $B$ for different  $k_0$ (Fig.~\ref{ColorMapsLin}(b)).
Similar color maps are presented  in Fig.~\ref{ColorMapsCirc} for the initial circular polarization with $S_{z0} = 1$, $S_{x,y0} = 0$.
The inclined broad-dashed (left) lines in Figs.~\ref{ColorMapsLin}(a) and~\ref{ColorMapsCirc}(a) indicate the turning times, $t _{1} = k_{y0} / \beta$, when polaritons reach the furthest point in $y$ direction for the given  $k_0$.
At this moment $k_{y}(t _{1}) = 0$.
The inclined narrow-dashed (right) lines indicate the time $t _{2} = 2 t_{1}$ when polaritons return to the starting point $\mathbf{r}_{c} (t_2)= (0,0)$.
The horizontal dash-dotted lines label the polarization state dynamics at $k_0 = 3.4 \, \mu \text{m} ^{-1}$.
The vertical dashed lines in Figs.~\ref{ColorMapsLin}(b) and~\ref{ColorMapsCirc}(b) correspond to~$t _{1}$ (left) and $t _{2}$ (right).

Let us now consider in detail the main peculiarities of the polariton polarization dynamics illustrated by Figs.~\ref{ColorMapsLin}--\ref{ColorMapsCirc}.
First, the central colorless horizontal stripes in the left panels in Figs.~\ref{ColorMapsLin}(b) and~\ref{ColorMapsCirc}(b) corresponding to $B \approx 0 $ confirm that $S_{x}$ remains unchanged during the propagation in the absence of an external applied magnetic field.

Second, for the strong enough magnetic field, $B \gg 1$~T where $\Omega _{cz}$ sufficiently exceeds $\Omega _{cx, cy}$, the oscillation frequency $\Omega_c$ tends to a constant value, $\Omega_c \rightarrow \Omega _{cz} = const$.
For the cases of both linear and circular initial polarizations and a large magnetic field, $S_{x}$ polarization component decreases at long times, see the left lower panels in Figs.~\ref{ColorMapsLin}(b) and~\ref{ColorMapsCirc}(b).
This can be understood in terms of the interplay between the dynamics of the effective magnetic field and the dynamics of of the polariton Stokes vector that precesses around this field.

Even stronger the interplay dynamical effects manifest themselves in the intermediate regime of $\Omega _{cz} \sim \Omega _{cx, cy}$.
This regime is characterized by the enhancement of the $S_{x}$ component of the Stokes vector accompanied by suppression of the other components, $S_{y}$ and $S_{z}$.
The latter is visualised as pale stripes in Figs.~\ref{ColorMapsLin} and~\ref{ColorMapsCirc}.
Herewith, for the given $B$, there is a discrete spectrum of $k$ that corresponds to the appearance of such stripes.
The value of $k$ increases with the increase of the index of the stripe $n$ approximately quadratically $k_n ^2 \varpropto n$.
Having in mind that $\Omega _{cx} \varpropto k^2$, this is a signature of a the resonant parametric character of the formation of closed trajectories at the surface of the Poincar\'{e} sphere as will be discussed below.

\begin{figure}[t]
\includegraphics[width= 1.\textwidth]{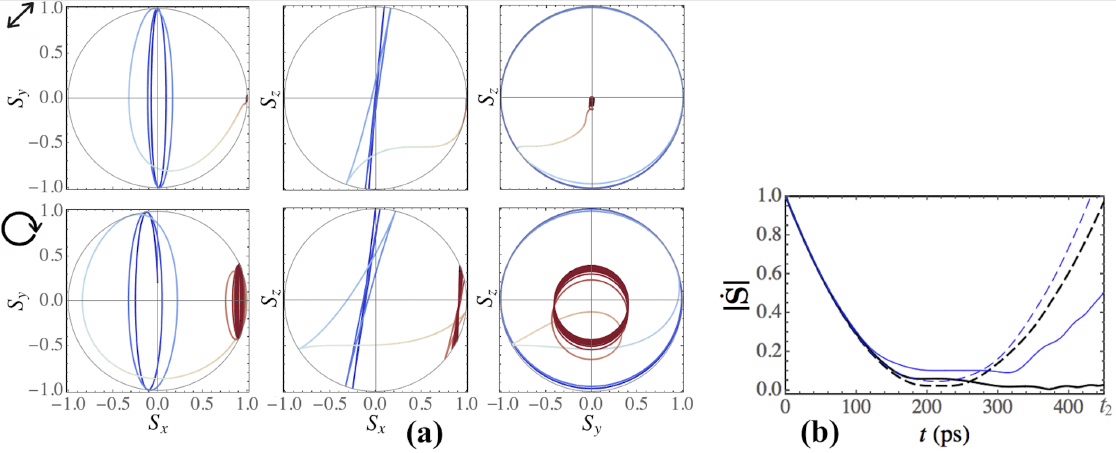}
\caption{\label{Attractors} (Color online)
(a)~Projections of the Poincar\'{e} sphere and the trajectories of the polariton Stokes vector.
In the upper panels $B=1.7$~T and the initial polarization is taken diagonal linear with the initial conditions $S_{y0} = 1$, $S_{x,z0} = 0$.
In the lower panels $B=2.8$~T and the initial polarization is taken circular with the initial conditions $S_{z0} = 1$, $S_{x,y0} = 0$.
The blue-colored parts of the curves show the Stokes vector trajectories from $t=0$ to $t_{1}$.
The red-colored curves show the Stokes vector evolution after $t_{1}$.
(b)~The absolute value of the time derivative of the polariton Stokes vector calculated as a function of time.
The dependence is shown in the interval $0\le t \le t_{2}$.
The value of $\dot{\mathbf{S}} (t)$ is normalized to $\dot{\mathbf{S}} (0)$.
The bold black curves correspond to the diagonal linear initial polarization with $S_{y0} = 1$ while the thin blue curves correspond to the circular initial polarization with $S_{z0} = 1$.
The initial wave number is taken $k_0 = 3.4 \, \mu \text{m}^{-1}$.
The values of $B$ are taken $0.8$~T (black dashed curve), $1.7$~T (black solid),
$2.8$~T (blue solid) and $1.4$~T (blue dashed).
}
\end{figure}

To analyse principal peculiarities of the discovered non-trivial polarization dynamics, let us track the corresponding phase trajectories on the Poincar\'{e} sphere.
Examples of the resulting trajectories are shown in Fig.~\ref{Attractors}(a) for the linear (${S_{y0} = 1}$, upper panels) and circular (${S_{z0} = 1}$, lower panels) initial polarizations.
The considered states are characterized by low-amplitude oscillations $s_{x,y,z}(t)$ of the polarization parameters at $t \rightarrow \infty$ around some stationary point (attractor) with $S_{x}(\infty)$ different from zero. 
We represent the Stokes parameters as $S_{x,y,z} (t) = \langle S_{x,y,z} (t) \rangle + s_{x,y,z}(t)$, where
$\langle S_{x,y,z} (t) \rangle$ characterize slowly changing mean values, $| \langle S_{x,y,z} \rangle| \gg |s_{x,y,z}|$ and $\partial _{t} \langle S_{x,y,z} \rangle  \ll \partial _t s_{x,y,z}$.
Substituting these expansions in~\eqref{StokesCompOnTime}, we finally obtain that $ \langle S_{y} \rangle = 0$ while the rest parameters are related  between themselves as $\langle S_{z} \rangle = \frac{\Omega_{cz}}{\Omega_{cx}} \langle S_{x}\rangle$.
Since $\Omega_{cx}$ grows with time as $t^2$ while $\Omega _{cz}$ remains constant, for large enough time $\langle S_{z} \rangle$ tends to zero.
Whereas the parameters $S_{x,y,z}$ are linked with each other trough $S_0$, the $ \langle S_{x} \rangle$ polarization component tends to 1 for $t \rightarrow \infty$.
Consequently, the discovered closed trajectories on the Poincar\'{e} sphere correspond to the oscillations of the Stokes vector in the $S_{y}S_{z}$ plane around the point $(\langle S_{x} \rangle, \langle S_{y} \rangle, \langle S_{z} \rangle)$ =$(0,0,0)$, see Fig.~\ref{Attractors}(a).
Thereby, applying the external magnetic field of a specific magnitude for a given incidence angle, one can achieve switching from the diagonal linear polarization state ($S_{y}=\pm 1$) or the circular polarization state ($S_{z}=\pm 1$) to the linear polarization state ($S_{x} \simeq \pm 1$).

Let us consider the behavior of the absolute value of the vector $\dot{ \mathbf{S} }$ given by Eq.~\eqref{StokesCompOnTime}.
This vector characterizes precession of the polariton Stokes vector around the effective magnetic field $\mathbf{\Omega} = (\Omega_{cx}, \Omega_{cy}, \Omega_{cz})$.
For the considered case where $\Omega_{cy} = 0$, Eq.~\eqref{StokesCompOnTime} yields
\begin{equation}
|\dot{ \mathbf{S} }|^2 = \Omega _{cx} ^2 \left(S_{y}^2 + S_{z}^2 \right) + \Omega _{cz} ^2 \left(S_{x}^2 + S_{y}^2 \right) - 2 \Omega _{cx} \Omega _{cz} S_x S_z.
\end{equation}

The temporal behavior of $|\dot{ \mathbf{S} }|$ is illustrated in Fig.~\ref{Attractors}(b) for a number of particular cases.
It is clearly seen that for the parameters corresponding to the pronounced polarization switch trajectories in Fig.~\ref{Attractors}(a) there is a significant reduction of $|\dot{ \mathbf{S} }|$ (solid lines in Fig.~\ref{Attractors}(b)) at the times around $t _{2}$.
In the limiting case when $|\dot{ \mathbf{S} }|$ tends to zero at large time, the trajectories degenerate to the point.
The closest to this regime trajectories are shown in the upper panels of Fig.~\ref{Attractors}(a) while the corresponding dynamics of $|\dot{ \mathbf{S} }|$ is shown in Fig.~\ref{Attractors}(b) (bold black solid curve).
The exact solutions of the equation $|\dot{ \mathbf{S} }| = 0$ are inachievable as $S_{x,y}$ change with time.
Nevertheless, if this equation is satisfied even approximately for $t \sim t_{2}$, the closed trajectories (attractors) may be seen on the surface of the Poincar\'{e} sphere.
$|\dot{ \mathbf{S} }|$ slowly increases with time at $t \rightarrow t_{2}$ in this regime.

\subsection*{Self-Interference of Polarized Polariton Waves}


Let us now consider the different excitation regime, namely the resonant cw excitation of polaritons in the spot characterized by $\mathbf{r}_c (0) = (0,0)$ and $k_{x0} = 0$, $k_{y0} = k_0$.
Obviously, in this case the polariton polarization dynamics is enriched by the effects coming from the self-interference of the polariton wave.
Since the pseudospin formalism is not applicable for the description of the interference effects, here we deal directly with the polariton field amplitudes~$\Psi_{\pm}$ governed by Eqs.~\eqref{ShiftedWFDef}--\eqref{GlobalPhaseDef}.

\begin{figure}[t]
\includegraphics[width= 1.\textwidth]{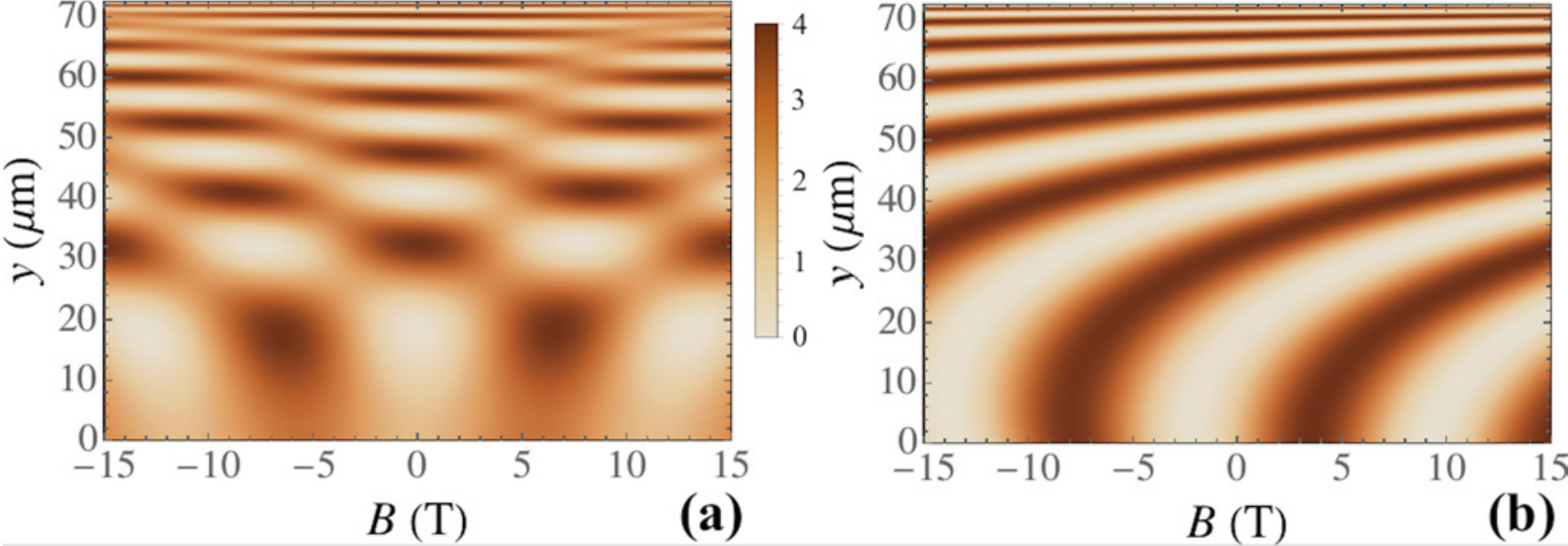}
\caption{\label{SelfInterf} (Color online) 
Patterns resulting from the self-interferense of the polariton waves propagating along the $y$ axis as functions of the magnitude of the external magnetic field~$B$.
(a) corresponds to the diagonal linear initial polarization, $S_{y0}=1$, $S_{x,z0}=0$,
(b) corresponds to the circullar initial polarization, $S_{z0}=1$, $S_{x,y0}=0$.
The initial wave number is taken $k_0 = 1 \, \mu \text{m}^{-1}$. 
}
\end{figure}

Figure~\ref{SelfInterf} shows the interference intensity patterns as functions of $B$ for the linear, $S_{y0}=1$ and $S_{x,z0} =0$, (a) and the circular, $S_{z0}=1$ and $S_{x,y0} =0$, (b) initial polarizations.
The pictures are rather different from the interference patterns of an incident and a reflected waves in the case of a conventional coherent reflection of light by a mirror.
They mostly demonstrate the Airy-like profile in $y$ direction.
In addition, in contrast with the conventional reflection by the mirror, the phase of the backward wave changes continuously and does not show a discontinuity with the incident wave phase at the mirror surface.
This is a signature of the ``slow mirror'' effect~\cite{Optica21}.

In the particular case corresponding to the purely polarized initial wave, e.g. the circularly polarized wave considered in Fig.~\ref{SelfInterf}(b), and in the absence of the magnetic field, $B=0$, the interference picture can be described analytically.
To do so, we formally split the ballistically propagating polariton wave into two waves, namely the forward wave propagating ``uphill'' (toward higher potential energies and narrower cavity widths)
 $\mathbf{\Psi} ^{(\text{up})}$  and the backward wave propagating ``downhill'' (in the opposite direction) $\mathbf{\Psi} ^{(\text{dn})}$ denoting $\mathbf{\Psi}^{(\text{up,dn})} = ( \Psi _{+}^{(\text{up,dn})}, \Psi _{-}^{(\text{up,dn})} ) ^{\text{T}}$.
It is convenient to introduce the ``time of flight'' parameter $\tau \equiv \tau (y)$ being a time of propagation of light emitted by a source at $y=0$ until the point having a coordinate $y$.
Note that light passes twice through each point except those with $y>y_{\text{max}}$, where $y_{\text{max}}$ is given by $y_{\text{max}} = \left. \hbar k_{y0}^2 \right/ 2 \beta m^{*}$. 
But we mean the shortest time of flight from the source to the point $y$.

To consider $\mathbf{\Psi} ^{(\text{up})}$  and $\mathbf{\Psi} ^{(\text{dn})}$ as co-propagating waves,
we shift the origin of the $t$ axis to $t_1$ and make the time reversion, $t \rightarrow -t$, i.e. we re-introduce the backward wave as $\mathbf{\Psi}^{(\text{dn})}(t) \rightarrow \mathbf{\Psi}^{(\text{dn})}( t_{1} - t)$.
Starting with the solutions $\Psi_{+} = \cos \left[ \kappa(t)/2 \right] \exp[i \chi (t)]$ and $\Psi_{-} = i \sin \left[ \kappa(t)/2 \right] \exp[i \chi (t)]$ with $\kappa(t)$  given above we finally obtain the resulting interference pattern as
\begin{equation}
\label{AnalytInterf1D}
I_{y>0} = 2 +
\cos \left[ k_{cy} (\tau) d_p (\tau) \right] +
\cos \left[ k_{cy} (\tau) d_m (\tau) \right],
\end{equation}
where the effective coordinates $ d_{p,m} (\tau)$ are given by
$$
d_{p,m} (\tau) = - \left[ \left( k_{cy}(\tau) \right)^2 \left(2 m^{*} \Delta _{\text{LT}} \mp \hbar \right) \pm 3 \hbar k_{y0}^2 \right] / 6 m^{*} \beta,
$$
and the time of flight parameter changes from $\tau=0$ to $t _1$ to cover the distance from $y=0$ to  $y_{\text{max}}$.
The expression~\eqref{AnalytInterf1D} describes the trajectory in Fig.~\ref{SelfInterf}(b) at $B=0$.

In the presence of the external magnetic field the interference patterns produced by self-intefering polaritons can not be described by a simple analytical expression.
The magnetic field effect on the interference patterns is presented in Fig.~\ref{SelfInterf}.
It is obvious from the figures that the magnetic field not only introduces a phase shift to the interfering waves but also affects the spreading of the waves in space.
One should also mention that the patterns in Fig.~\ref{SelfInterf}(b) are asymmetric with respect to $B=0$, and the interference picture is reversed in the case of the opposite initial polarization, $S_{z0}=-1$.

\section*{Propagation of Polaritons in the Oblique to the Gradient Direction}

\subsection*{Single Polariton Propagation}

In practice, we have one more external control parameter in our problem that is the shooting angle $\theta$, the angle between the initial polariton wave vector and $x$-axis.
In this case, $k_x$ and the related parameter $\Omega _{cy}$ differ from zero.
The main difference from the case considered above is that now the particle propagates along a parabolic trajectory, hence the forward and the backward branches are separated by a distance $\frac{2 \hbar ^2 k^2}{m^{*} \beta} \frac{| \tan[\theta] |}{1+\tan^2 [\theta]}$ in the $x$ direction at the latitude $x = 0$.
In~\cite{Optica21} the spatial separation between forward and backward going polaritons helped measuring the long-living polariton lifetime since it allowed eliminating the spurious radiation resulting from the reflection of the excitation laser.

\begin{figure}[t]
\includegraphics[width= 1.\textwidth]{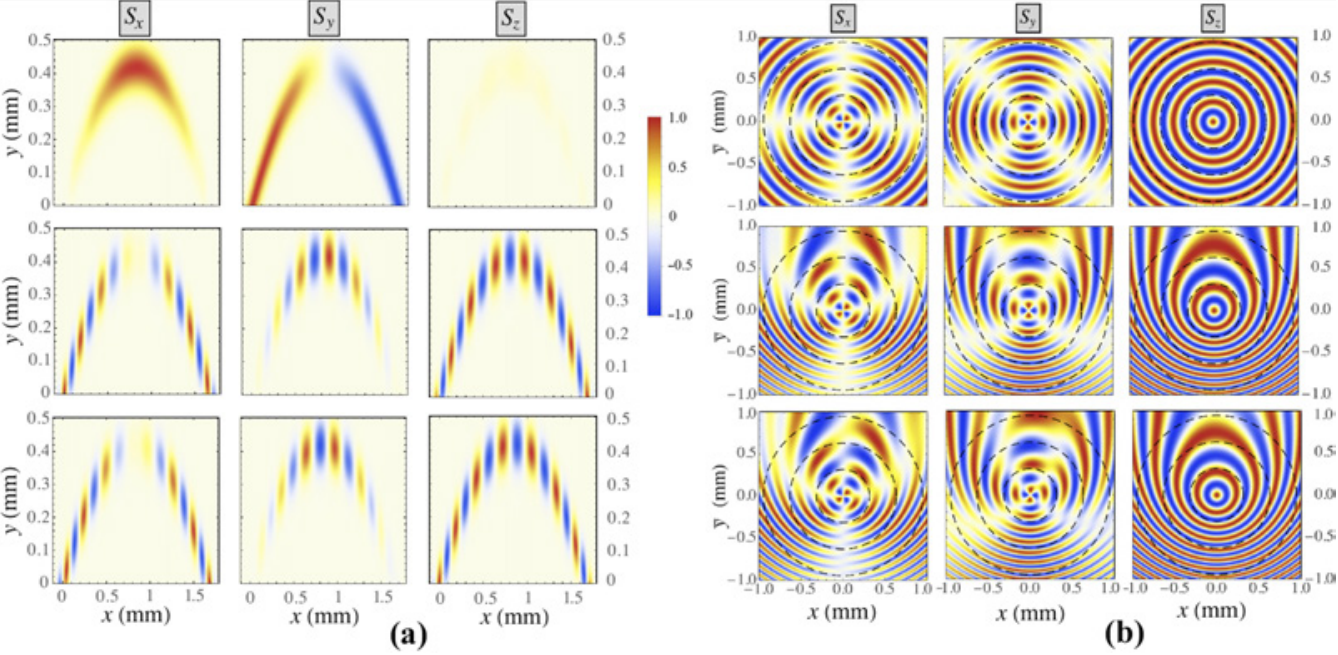}
\caption{\label{ObliquePropFigs} (Color online)
(a) Polariton polarization components $S_{x,y,z}$ of ballistically propagating exciton-polaritons at an oblique angle.
The initial polarization is taken diagonal linear with $S_{y0}=1$ for the upper panels,
linear with $S_{x0}=1$ for the middle panels and
circular with $S_{z0}=1$ for the lower panels.
The magnetic field $B=0$, the initial wave number is taken $k_0 = 3.4 \, \mu \text{m}^{-1}$, the shooting angle is taken $45 ^{\circ}$ ($k_{x0} = k_{y0} = \left. k \right/ \sqrt{2}$) for all the panels.
(b) Polarization patterns resulting from the pulsed excitation of polaritons by a point-like source calculated at different values of $\beta$ and $B$.
For the upper panels, $\beta = 0$, $B=0$;
for the middle panels, $\hbar \beta = 10.5 \, \mu \text{eV} / \mu \text{m} $, $B=0$;
for the lower panels, $\hbar \beta = 10.5 \, \mu \text{eV} / \mu \text{m}$, $B=3$~T.
The initial polarization is taken circular with the initial conditions $S_{x,y0} = 0$, $S_{z0} = 1$.
For the clarity of representation, in the middle and lower panels  we introduces a new dynamically shifted coordinate $\bar{y} = y \left. + \hbar \beta t^2 \right/ 2 m^{*}$.
Dashed circles (from the inner to the outer) correspond to $t = 40$~ps,  $t = 80$~ps and $t = 120$~ps.
}
\end{figure}

Figure~\ref{ObliquePropFigs}(a) demonstrates the spatial distribution of the polariton polarization components $S_{x,y,z}$ in the cavity plane for different initial conditions in the absence of the external magnetic field, $B=0$.
The initial polarizations are taken $S_{y0}=1$ for the upper panels, $S_{x0}=1$ for the middle panels and $S_{z0}=1$ for the lower panels.
The other initial polarization components are taken equal to zero in all cases.
The polarization components $S_{x,y,z}$ are multiplied by the spatially dependent wave packet intensity of the polariton field.
The incident wave packet is governed by the shape of the Gaussian form $E \sim \exp [ \left. -(y-y_{c})^{2} \right/ 2y_{w} ^{2}]$ with $y_{w} = 0.04$~mm and the time-dependent ``center of mass'' coordinates $\left( x_{c}(t), y_{c} (t) \right)$.
The upper panels in Fig.~\ref{ObliquePropFigs}(a) demonstrate the transformation of the diagonal linear polarization $S_{y0}=1$ to the antidiagonal one, $S_{y} = -1$.
The Stokes vector evolves from $S_{y0} = 1$ to $S_{y} = -1$ passing by the  $S_{x} = 1$ state.
The component $S_{z}$ is close to zero in this case.
The middle and the lower panels demonstrate the inversion of the linear (middle) and circular (lower) polarizations, as one can see comparering the polarization patterns before and after the turning point.

\subsection*{The Pulsed Excitation of Exciton-Polaritons by a Point-like Source}

We now consider the polarization dynamics of the cylindrical polariton wave packet emitted by a point-like source placed in the point $(0,0)$.
The initial polarization of the polaritons is taken circular with $S_{x0}=S_{y0} = 0$ and $S_{z0}=1$.
Figures~\ref{ObliquePropFigs}(b) represent the parametric dependencies of the polariton polarization components on the spatial coordinates: $x_{0}$ and the shifted coordinate $\bar{y} = y + \frac{ \hbar \beta t ^2 }{ 2 m^{*}}$.
The black dashed circles in Fig.~\ref{ObliquePropFigs}(b) correspond to $t=40$~ps, 80~ps and 120~ps (counting from the inner to the outer circle).

The upper row panels illustrate the trivial case without the ``gravitational'' force ($\beta = 0$) and the magnetic field ($B = 0$).
It is the case discussed in~Ref.~\cite{PRL109036404} and showing the signature of the optical spin Hall effect.
In the middle row panels, the ``gravitation'' is switched on, $\hbar \beta = 10.5 \, \mu\text{eV}/\mu\text{m}$.
In this case, the polarization patterns ``fall down'' in time and the symmetry (or anti-symmetry) of the patterns  with respect to the point moving with the velocity $ - \left. \hbar \beta t \right/ m^{*}$ down the $y$ direction is conserved.
In the lower row panels, the magnetic field $B=3$~T has been added as well.
The magnetic field $B$ breaks the pattern symmetry and induces the rotation of the patterns clockwise for $B<0$ and anticlockwise for $B>0$. 
Since we do not consider a continuous pump here, the interference effects are absent.

\subsection*{Self-Interference of a Polariton Flow Excited by a Point-like Source in the CW Regime}

\begin{figure}[t]
\includegraphics[width= 1.\textwidth]{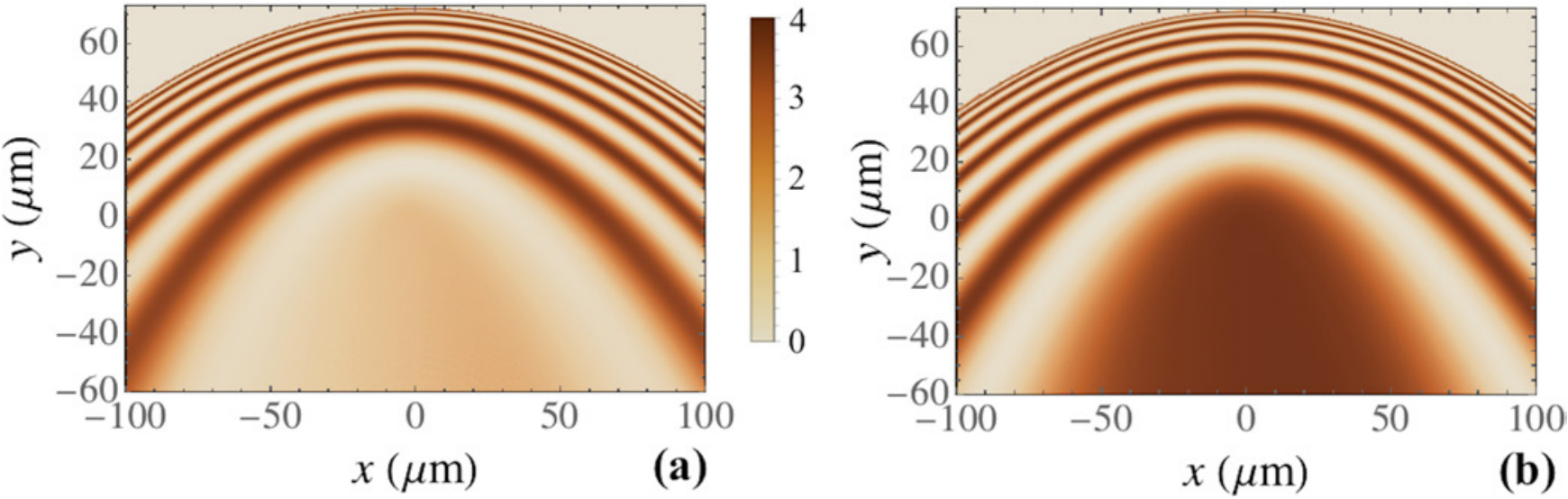}
\caption{\label{PointSelfInt} (Color online)
Patterns resulting from the self-interference of the circularly polarized polariton field emitted by a point source placed in $(0,0)$ point at~$B=0$ (a) and $B=3$~T (b).
The initial wave number is taken $k_0 = 1 \, \mu \text{m} ^{-1}$.
}
\end{figure}

In contrast with the previous subsection, here we consider the case where the point $(0,0)$ is a source of continuous polariton waves.
Figure~\ref{PointSelfInt} demonstrates the real-space intensity patterns appearing as a result of the self-interference of the polariton wave initially emitted as a cylindrical wave.
The main peculiarity of the patterns is that they are limited in $+y$ direction.
The ``slow mirror'' boundary corresponds to the envelope of the classical center-of-mass trajectories of the polaritons emitted by the point-like source.
The boundary possesses a parabolic shape obeying the equation
\begin{equation}
y_{\text{SM}} = \frac{\hbar k^2}{2 \beta m} - \frac{\beta m}{\hbar  k^2} x^2 .
\end{equation}

Another peculiarity is that the interference fringes are concentrated in the upper part of the interference picture, while the lower part is characterized by the uniform intensity distribution.
Above we obtained the analytical expression~\eqref{AnalytInterf1D} for the interference fringes at $y>0$ for the initially circullarly polarized polariton wave propagating along the gradient.
Now we carry out a similar procedure for $y<0$.

Let us first distinguish between the interfering polariton waves.
The first wave which we shall refer to as the forward wave for clarity is one emitted downhill with the wave vector $\mathbf{k}_0^{\text{dn}} = (0, -k_0)$.
Another wave that we refer to as the reflected wave is one emitted uphill with the wave vector $\mathbf{k}_0^{\text{up}} = (0, k_0)$, then reflected by the ``slow mirror'' and returned to the starting point after $t = t_2$ with the geometric phase shift.
Based on the known solutions we find that at $y<0$ the total intensity is independent on the coordinate $y$: $I_{y<0} = 2+ \cos \left[ \left. k_0 ^3 (\Delta _{\text{LT}} + \hbar / m^{*}) \right/ 3 \beta\right] +\cos \left[ \left. k_0 ^3 (\Delta _{\text{LT}} - \hbar / m^{*}) \right/ 3 \beta\right]$.

The effect of the external magnetic field on the interference patterns is demonstrated in Fig.~\ref{PointSelfInt}(b).
It introduces a phase shift in the interfering waves that results in the shift of the intensity fringes in the upper part of the interference picture and in the change of the total intensity in the lower part. 

\section*{Conclusions}

We have considered the dynamics of exciton-polaritons propagating over large distances in wedged microcavities in the presence of the TE-TM splitting and external magnetic fields.
The cavity thickness gradient playing role of the artificial gravity strongly affects the TE-TM splitting and the speed of propagating polaritons.
We demonstrated theoretically the formation of polarization patterns in the real space due to the optical spin Hall effect and ``slow reflection''.

We have demonstrated the possibility of manipulating the polariton polarization state by choosing the excitation laser incidence angle and the magnitude of the  magnetic field.
Interestingly at the specific combinations of the  magnetic field and the initial polariton wave vector the polariton Stokes vector tends to an attractor on the surface of the Poincar\'{e} sphere.

The ``gravitational'' force compelling polaritons to change both magnitude and direction of their wave vector is the cause of another fascinating effect that is the self-interference of a spin-polarized polariton field.
We show that for a point-like polariton source in the wedged microcavity the ``slow mirror'' has a parabolic shape determined by the ``gravitational'' force and the initial polariton wave number.
The effect of the external magnetic field on the interference patterns can be considered as well.
The effects discussed here may be used in future polariton spin transistors and logical gates.

\section*{Acknowledgments}
This work was supported by the EPSRC Programme grant on Hybrid Polaritonics No. EP/M025330/1.
The work of E.S.S. was supported by the Russian Foundation for Basic Research Grants No. 16-32-60104, No. 15-59-30406, and by grant of President of Russian Federation for state support of young Russian
scientists No. MK-8031.2016.2. 
A.V.K. acknowledges the partial support from the HORIZON 2020 RISE project CoExAn (Grant No. 644076).
E.S.S. acknowledges Dr. I.~Yu.~Chestnov for valuable advices and fruitful discussions.

\section*{Contributions}
A.V.K. initiated the study and proposed the theoretical model.
E.S.S. conducted the theoretical simulations and wrote the manuscript.

\section*{Competing interests}
The authors declare no competing financial interests.

\end{document}